\documentclass[twocolumn,prx,longbibliography,nofootinbib, superscriptaddress]{revtex4-2}
\usepackage[cal=cm]{mathalfa}
\usepackage{graphicx} 
\usepackage{amsfonts}
\usepackage{amssymb}
\usepackage{amscd}
\usepackage{amsmath}    
\usepackage{amsthm}
\usepackage{appendix}
\usepackage{mathptmx}
\usepackage{enumerate}
\usepackage{enumitem}
\usepackage{upgreek}
\usepackage{epsfig}
\usepackage{float}
\usepackage{xcolor}
\definecolor{LinkBlue}{HTML}{1F4E79}
\definecolor{AcademicBlue}{HTML}{24527A}
\newif\ifrevision
\revisiontrue
\definecolor{RevisionBlue}{RGB}{0,65,190}

\usepackage[
  colorlinks=true,
  linkcolor=AcademicBlue,
  citecolor=AcademicBlue,
  urlcolor=AcademicBlue
]{hyperref}
\usepackage{cleveref}

\crefname{theorem}{theorem}{theorems}
\Crefname{theorem}{Theorem}{Theorems}

\crefmultiformat{theorem}
  {theorems~#2#1#3}
  {,~#2#1#3}
  {,~#2#1#3}
  {,~#2#1#3}

\Crefmultiformat{theorem}
  {Theorems~#2#1#3}
  {,~#2#1#3}
  {,~#2#1#3}
  {,~#2#1#3}

\usepackage{makecell}
\usepackage{times}
\usepackage{pifont}
\usepackage{subfigure}
\usepackage{subfloat}
\usepackage{xcolor}		
\usepackage{physics}
\usepackage[most]{tcolorbox}
\usepackage{tabularx}
\usepackage{booktabs}
\usepackage{amssymb}
\usepackage{wasysym}
\usepackage{tikz}
\usepackage{tikzpeople}
\usepackage{pgfplots}
\usepackage{pifont}
\usetikzlibrary{quantikz}
\usetikzlibrary{arrows}
\usetikzlibrary{shapes,fadings,snakes}
\usetikzlibrary{decorations.pathmorphing,patterns}
\usetikzlibrary{calc}
\usetikzlibrary{positioning}

\newtheorem{theorem}{Theorem}

\DeclareDocumentCommand\dket{ s m }{%
    \IfBooleanTF{#1}
    {\lvert #2\rangle\!\rangle}
    {\left\lvert #2\middle\rangle\!\middle\rangle\right.}
}

\DeclareDocumentCommand\dbra{ s m }{%
    \IfBooleanTF{#1}
    {\langle\!\langle #2\rvert}
    {\left.\middle\langle\!\middle\langle #2\right\rvert}
}

\DeclareDocumentCommand\dketbra{ s m g }{%
    \IfBooleanTF{#1}
    {%
        \IfNoValueTF{#3}
        {\lvert #2\rangle\!\rangle\langle\!\langle #2\rvert}
        {\lvert #2\rangle\!\rangle\langle\!\langle #3\rvert}
    }
    {%
        \IfNoValueTF{#3}
        {\left\lvert #2
         \middle\rangle\!\middle\rangle
         \middle\langle\!\middle\langle #2
         \right\rvert}
        {\left\lvert #2
         \middle\rangle\!\middle\rangle
         \middle\langle\!\middle\langle #3
         \right\rvert}
    }
}

\DeclareDocumentCommand\dbraket{ s m m }{%
    \IfBooleanTF{#1}
    {\langle\!\langle #2\mid #3\rangle\!\rangle}
    {\left.
     \middle\langle\!\middle\langle #2
     \middle| #3
     \middle\rangle\!\middle\rangle
     \right.}
}

\usepackage[most]{tcolorbox}
\newtcolorbox[auto counter]{mybox}[2][]{
	enhanced,
	breakable,
	colback=blue!5!white,
	colframe=blue!75!black,
	fonttitle=\bfseries,
	title=Box \thetcbcounter: #2,#1
}

\makeatletter

\newif\ifappendixtocrecord
\appendixtocrecordfalse

\newcommand{\appendixtableofcontents}{%
  \section*{Contents of Appendices}%
  \setcounter{tocdepth}{3}%
  \@starttoc{atoc}%
  \appendixtocrecordtrue
}

\let\oldsection\section
\let\oldsubsection\subsection
\let\oldsubsubsection\subsubsection

\renewcommand{\section}{%
  \@ifstar{\app@sectionstar}{\@ifnextchar[{\app@sectionopt}{\app@sectionnoopt}}%
}

\newcommand{\app@sectionstar}[1]{%
  \oldsection*{#1}%
}

\newcommand{\app@sectionnoopt}[1]{%
  \oldsection{#1}%
  \ifappendixtocrecord
    \addcontentsline{atoc}{section}{\protect\numberline{\thesection}#1}%
  \fi
}

\def\app@sectionopt[#1]#2{%
  \oldsection[#1]{#2}%
  \ifappendixtocrecord
    \addcontentsline{atoc}{section}{\protect\numberline{\thesection}#1}%
  \fi
}

\renewcommand{\subsection}{%
  \@ifstar{\app@subsectionstar}{\@ifnextchar[{\app@subsectionopt}{\app@subsectionnoopt}}%
}

\newcommand{\app@subsectionstar}[1]{%
  \oldsubsection*{#1}%
}

\newcommand{\app@subsectionnoopt}[1]{%
  \oldsubsection{#1}%
  \ifappendixtocrecord
    \addcontentsline{atoc}{subsection}{\protect\numberline{\thesubsection}#1}%
  \fi
}

\def\app@subsectionopt[#1]#2{%
  \oldsubsection[#1]{#2}%
  \ifappendixtocrecord
    \addcontentsline{atoc}{subsection}{\protect\numberline{\thesubsection}#1}%
  \fi
}

\renewcommand{\subsubsection}{%
  \@ifstar{\app@subsubsectionstar}{\@ifnextchar[{\app@subsubsectionopt}{\app@subsubsectionnoopt}}%
}

\newcommand{\app@subsubsectionstar}[1]{%
  \oldsubsubsection*{#1}%
}

\newcommand{\app@subsubsectionnoopt}[1]{%
  \oldsubsubsection{#1}%
  \ifappendixtocrecord
    \addcontentsline{atoc}{subsubsection}{\protect\numberline{\thesubsubsection}#1}%
  \fi
}

\def\app@subsubsectionopt[#1]#2{%
  \oldsubsubsection[#1]{#2}%
  \ifappendixtocrecord
    \addcontentsline{atoc}{subsubsection}{\protect\numberline{\thesubsubsection}#1}%
  \fi
}

\makeatother

\begin{document}
\title{On the geometry, typicality, and detectability of quantum magic}

\author{Zhenhuan Liu}
\thanks{\href{mailto:qubithuan@gmail.com}{qubithuan@gmail.com}}
\affiliation{Quantum Research Center, Technology Innovation Institute (TII), Abu Dhabi, United Arab Emirates}

\author{Zi-Wen Liu}
\thanks{\href{mailto:zwliu0@tsinghua.edu.cn}{zwliu0@tsinghua.edu.cn}}
\affiliation{Yau Mathematical Sciences Center, Tsinghua University, Beijing 100084, China}

\begin{abstract}
Quantum magic, or nonstabilizerness, is a central resource for quantum computation, motivating a systematic study of its mathematical structure and physical behavior. Here we characterize the geometry of the stabilizer polytope, obtaining near-tight bounds on its inradius, volume radius, and number of facets. These results reveal a pronounced separation between the typical presence of magic and its detectability. In particular, random mixed states obtained by tracing out an environment remain magical with high probability even when the environment contains nearly twice as many qubits as the system, demonstrating parametrically greater robustness to environmental mixing than bipartite entanglement. In contrast, commonly used magic criteria, including the Triangle and Pauli-norm-based criteria, lose detection power under much weaker mixing. More generally, when the number of environment qubits scales proportionally with the system size, certification with constant success probability requires exponentially many linear observables, even given their exact expectation values. We also prove that the number of facets of the stabilizer polytope grows doubly exponentially with the number of qubits, making exhaustive witness enumeration impractical for complete magic certification and motivating approaches that exploit the structure and symmetries of stabilizer states.
\end{abstract}

\maketitle

\section{Introduction}

An important paradigm in quantum information science is to characterize certain nonclassical properties of quantum systems as operational
resources underlying information-processing advantages~\cite{chitambar2019quantum}.
Prominent examples include coherence, entanglement, and quantum magic, each characterized by a distinguished set of free states whose manipulation alone does not provide the corresponding resource.
For entanglement, the free states are separable states, namely convex mixtures of product states~\cite{Horodecki2009EntanglementRMP};
for quantum magic, they are convex mixtures of pure stabilizer states~\cite{Veitch2014ResourceTheoryStabilizer}.
Understanding the geometry of these free state sets is fundamental to characterizing the corresponding quantum resources and their operational properties.

Entanglement theory offers a paradigmatic example of the power of this geometric viewpoint.
Extensive studies of the separable set have characterized the size of separable balls around the maximally mixed state and the asymptotic volume of the set~\cite{Zyczkowski1998Volume,Gurvits2002largest,szarek2005volume,AubrunSzarek2006Volume}. 
These geometric properties have guided the design of entanglement detection protocols~\cite{GuhneToth2009EntanglementDetection,Barreiro2010multipartite} and revealed fundamental limits on entanglement detectability~\cite{liu2022detectability}.
The important questions of when entanglement typically survives environmental mixing and how its presence depends on the size of the environment have also been addressed using geometric methods~\cite{zyczkowski2001induced,aubrun2012phase,AubrunSzarekYe2014}.
Furthermore, related studies have shown how the separable structure of high-temperature Gibbs states can be exploited for efficient classical simulation and state preparation~\cite{Bakshi2024Gibbs}.
Together, these developments highlight the importance of geometric and structural information about a free set for understanding the operational, statistical, and computational properties of the corresponding resource.

Despite the central role of magic in universal quantum computation
~\cite{gottesman1998heisenberg,aaronson2004improved,bravyi2005universal},
the number of stabilizer states and the combinatorial complexity of
their convex hull grow rapidly with the number of qubits, making the
stabilizer polytope notoriously difficult to characterize.
Previous work has studied geometric relations and distances among
stabilizer states~\cite{garcia2017geometry}, structural properties of dual stabilizer polytopes~\cite{okay2021extremal,IpekOkay2023Degenerate,Zurel2024HiddenVariable}, and general bounds on magic in multiqubit states~\cite{liu2022many}.
More recently, the Triangle Criterion identified boundary states of
purity $1/(d-1/2)$ and conjectured that the resulting upper bound on
the Hilbert--Schmidt inradius is tight~\cite{LiuHaugYe2026Triangle}.
Subsequently, the inradius was determined exactly for odd-prime-dimensional qudits~\cite{zurel2026basis}.
At the combinatorial level, the known lower and upper bounds on the
number of facets remain widely separated
~\cite{Karanjai2018Contextuality,ZurelOkayRaussendorf2024Bits}.
Taken together, these results reveal rich structure of the stabilizer polytope, but leave its key geometric features far from well understood.

In this work, we quantitatively characterize the stabilizer polytope from three complementary perspectives: its local geometry, global geometry, and combinatorial complexity, and use them to derive a series of results concerning the typicality and detectability of quantum magic.
We first establish a lower bound on the Hilbert--Schmidt inradius of the \(n\)-qubit stabilizer polytope that nearly matches the conjectured inradius given by the Triangle Criterion~\cite{LiuHaugYe2026Triangle}. Combining this bound with a volume estimate based on the number of pure stabilizer states, we show that the volume radius scales as \(1/d\) up to a logarithmic factor, where \(d=2^n\) is the system dimension.

These geometric results reveal a substantial separation between the typical presence of magic and its detectability. Consider an \(n\)-qubit system and an \(m\)-qubit environment in a joint Haar-random pure state. As \(m\) increases, the reduced state of the system undergoes a sharp transition from typically magical to typically magic-free at roughly \(2n\) environment qubits. Entanglement across a fixed balanced bipartition of the system disappears earlier, at roughly \(3n/2\) environment qubits~\cite{AubrunSzarekYe2014}. These thresholds uncover a broad regime in which the system  typically has magic yet is separable. Despite this persistence, standard criteria for mixed-state magic including the Pauli norm and Triangle criteria~\cite{campbell2011catalysis,LiuHaugYe2026Triangle} already fail with high probability for much smaller environments. Our inradius bound establishes a limitation that extends beyond these particular criteria: even with access to exact expectation values, any protocol based on fixed linear observables requires \(\Omega(2^m)\) distinct observables to certify magic with nonvanishing probability. In particular, this indicates that as long as the environment is not vanishingly small compared to the system, an exponential number of observables is required. 

Finally, by combining the above radius results with convex duality, we show that the number of facets of the stabilizer polytope satisfies \(\log F_d=\widetilde{\Theta}(d^2)\). Consequently, a complete description of the polytope by linear magic witnesses requires doubly exponentially many inequalities. These results motivate exploiting the symmetries of stabilizer states to reduce the computational cost of magic detection and highlight the limitations of existing contextuality-based descriptions of the stabilizer polytope~\cite{leone2026unbearable,Karanjai2018Contextuality,ZurelOkayRaussendorf2024Bits}.

\section{Preliminaries}
\label{sec:preliminaries}

For an $n$-qubit system, let $d=2^n$ denote the Hilbert space dimension.
A pure state is called a stabilizer state when it is the common eigenstate of a maximal set of mutually commuting Pauli operators.
The set of mixed stabilizer states is the convex hull of all pure stabilizer states,
\begin{equation}
\operatorname{STAB}_n
=
\operatorname{conv}
\left\{
\ketbra{\phi}{\phi}:
\ket{\phi} \text{ is a stabilizer state}
\right\}.
\end{equation}
States outside $\operatorname{STAB}_n$ are referred to as magic states.
Since the number of pure stabilizer states is finite, $\operatorname{STAB}_n$ forms a convex polytope in the $(d^2-1)$-dimensional state space~\cite{garcia2017geometry}.

Throughout this work, it is convenient to center the stabilizer polytope around the maximally mixed state, $K_n:=\operatorname{STAB}_n-\mathbb{I}/d$.
Its Hilbert--Schmidt inradius $r_{\mathrm{in}}(K_n)$ is the radius of the largest Euclidean ball contained in $K_n$, which, by Pauli symmetry, can always be chosen to be centered at the origin.
Since $\|\rho-\mathbb{I}/d\|_2^2=\operatorname{Tr}(\rho^2)-1/d$, the radius of the largest origin-centered ball is directly equivalent to the purity threshold of magic states, and hence coincides with $r_{\mathrm{in}}(K_n)$.
We will also use the volume radius
\begin{equation}
\operatorname{vrad}(K)
=
\left(
\frac{\operatorname{Vol}(K)}
{\operatorname{Vol}(B_2^{d^2-1})}
\right)^{1/(d^2-1)},
\end{equation}
which is the radius of a Euclidean ball having the same volume as $K$ and provides a natural measure of the global size of the convex body.

For any full-dimensional convex body $K$ containing the origin in its interior, we denote its polar by $K^\circ:=\{Y:\langle X,Y\rangle\leq1,\ \forall X\in K\}$.
Polarity reverses inclusion and exchanges vertices and facets. In particular, the facets of $K_n$ are in one-to-one correspondence with the vertices of $K_n^\circ$.
Furthermore,  a landmark result of Bourgain and Milman in asymptotic convex geometry known as the reverse Santal\'o inequality relates the volumes of a centrally symmetric convex body $K$ and its polar $K^\circ$, asserting that the product of their volume radii satisfies $\operatorname{vrad}(K)\operatorname{vrad}(K^\circ)\geq c$ for a dimension-independent universal constant $c>0$~\cite{bourgain1987new}\footnote{Equivalently, its volume product, or Mahler volume, obeys
\begin{equation}
\operatorname{Vol}(K)\operatorname{Vol}(K^\circ)\geq c^m\operatorname{Vol}(B_2^m)^2.
\end{equation}}.
For a convex body $K$ containing the origin in the interior, its gauge is defined as
$\|X\|_K:=\inf\{t>0:X\in tK\}$, such that $X\in K$ if and only if $\|X\|_K\leq1$.

We use the standard asymptotic notation throughout the paper.
For positive functions $f$ and $g$, we write $f=\mathcal{O}(g)$ if $f/g$ is bounded above by a constant, $f=\Omega(g)$ if it is bounded below by a positive constant, and $f=\Theta(g)$ if both bounds hold.
The soft notation $f=\widetilde{\Theta}(g)$ means that both lower and upper bounds hold up to polylogarithmic factors.
Furthermore, $f=o(g)$ means $f/g\to0$ and $f\sim g$ means $f/g\to1$ in the relevant asymptotic limit.
All implicit constants are independent of the system size unless stated otherwise.

\section{Radii of Stabilizer Polytopes}
\label{sec:radius}

We begin by studying the local geometric property of the stabilizer polytope, namely its purity threshold and inradius.

\begin{theorem}[Purity threshold and inradius bounds]
\label{thm:purity-radius}
Let \(n\ge 1\), let \(d=2^n\), and define
\begin{equation}
    a_\ast \coloneqq \frac{1}{2.181847} =
    0.458327\cdots.
\end{equation}
Then every \(n\)-qubit density operator \(\rho\) satisfying
\begin{equation}
    \operatorname{Tr}(\rho^2)
    \le
    \frac{1}{d-a_\ast}
\end{equation}
belongs to the stabilizer polytope $\operatorname{STAB}_n$.
Hence, the centered stabilizer polytope $K_n\coloneqq\operatorname{STAB}_n-\frac{\mathbb I}{d}$ has Hilbert--Schmidt inradius satisfying
\begin{equation}
\sqrt{\frac{a_\ast}{d(d-a_\ast)}}\le r_{\mathrm{in}}(K_n)\le \sqrt{\frac{0.5}{d(d-0.5)}}.
\end{equation}
\end{theorem}

The proof of the lower bound, detailed in Appendix~\ref{app:purity-radius}, proceeds in the polar geometry of the stabilizer polytope.
We use the stabilizer-positive dual representation, namely the set of trace-one Hermitian operators having nonnegative overlap with every pure stabilizer state, and derive a dimension-doubling recursion for its maximal Hilbert--Schmidt norm.
Combined with a sharp spectral bound on the third moment, the exact two-qubit solution~\cite{reichardt2006quantum}, and a finite rational certificate, this yields the dimension-independent bound required for Theorem~\ref{thm:purity-radius}.
The upper bound is given by Triangle Criterion, which provides a complementary construction showing that magic can occur arbitrarily close to purity $1/(d-1/2)$, corresponding to the upper bound $r_{\mathrm{in}}(K_n)\leq 1/\sqrt{d(2d-1)}$~\cite{LiuHaugYe2026Triangle}.

Although our lower bound does not close the remaining constant gap between $a_\ast=0.458327\cdots$ and $1/2$, this gap is independent of the number of qubits.
It is also instructive to compare this behavior with the odd-prime-dimensional qudit case, where the  purity threshold has been determined exactly as $1/(d-1/d)$, corresponding to an inradius $1/\sqrt{d(d^2-1)}$~\cite{zurel2026basis}.
Thus, in the sense captured by the inradius, qubit stabilizer states provide a substantially more even coverage around the maximally mixed state than their odd-prime-dimensional qudit counterparts.
As we show below, this dimension-independent control is sufficient to obtain strong consequences for the global and combinatorial geometry of the stabilizer polytope.

We next turn from the closest boundary point to the global size of the stabilizer polytope.
Recall that $\operatorname{vrad}(K_n)$ denotes the radius of a Euclidean ball having the same $(d^2-1)$-dimensional volume as $K_n$.

\begin{theorem}[Volume radius]
\label{thm:volume-radius}
Let $d=2^n$. The volume radius of $n$-qubit centered stabilizer polytope satisfies
\begin{equation}
\Omega\left(\frac{1}{d}\right)
\leq
\operatorname{vrad}(K_n)
\leq
\mathcal{O}\!\left(\frac{\log d}{d}\right).
\end{equation}
\end{theorem}

The detailed proof is deferred to Appendix~\ref{app:volume-radius}, and we outline the core idea here.
The lower bound follows immediately from Theorem~\ref{thm:purity-radius}, since a convex body has volume at least that of its inscribed Euclidean ball.
For the upper bound, we use a general result in high-dimensional convex geometry~\cite{carl1988gelfand,barany1988approximation}: an $m$-dimensional polytope contained in the Euclidean unit ball and having $N$ vertices has volume radius at most $C\sqrt{\log(eN/m)/m}$.
The stabilizer polytope has dimension $m=d^2-1$ and $N_n=d\prod_{j=1}^n(2^j+1)$ vertices, for which $\log N_n=\Theta(n^2)$~\cite{garcia2017geometry}.
The Carl--Pajor bound therefore gives $\operatorname{vrad}(K_n)=\mathcal{O}(n/d)$.
Consequently, the volume radius of the stabilizer polytope is determined up to only a factor $\mathcal{O}(n)=\mathcal{O}(\log d)$.

It is instructive to compare these geometric scales with those of bipartite entanglement.
For a balanced bipartite system of total Hilbert space dimension $d$, the separable set has the same Hilbert--Schmidt inradius scaling, $r_{\mathrm{in}}=\Theta(d^{-1})$~\cite{Gurvits2002largest}, but its volume radius scales as $\Theta(d^{-3/4})$~\cite{AubrunSzarek2006Volume}.
In contrast, for the stabilizer polytope these two geometric scales differ by at most a logarithmic factor.
This reveals a qualitative distinction in global geometry: the separable set extends parametrically farther beyond its largest inscribed ball, whereas the local and global scales of the stabilizer polytope remain much more closely matched.

\section{Magic Typicality}
\label{sec:typicality}

Given these geometric properties, we now turn to the typical behavior of magic in random mixed states.
For integers $d$ and $k$, let $\pi_{d,k}$ denote the induced probability distribution obtained by drawing a Haar-random pure state $\ket{\Psi}$ on $\mathbb{C}^{d}\otimes\mathbb{C}^{k}$ and tracing out the second subsystem,
\begin{equation}
\rho_{d,k}
=
\operatorname{Tr}_{\mathbb{C}^{k}}
\ketbra{\Psi}{\Psi},
\qquad
\rho_{d,k}\sim\pi_{d,k}.
\label{eq:induced-state}
\end{equation}
The parameter $k$ can be viewed as the dimension of an inaccessible environment, and increasing $k$ drives the reduced state toward the maximally mixed state~\cite{zyczkowski2001induced,AubrunSzarekYe2014}.
We say that a sequence $k_\star(d)$ is a \emph{magic typicality threshold} if, for every fixed $\epsilon>0$, the probability of magic approaches one below the threshold and zero above it, namely,
\begin{equation}
\begin{aligned}
k\leq(1-\epsilon)k_\star(d)
&\quad\Longrightarrow\quad
\Pr_{\rho\sim\pi_{d,k}}
[\rho\notin\operatorname{STAB}_n]\longrightarrow1,\\
k\geq(1+\epsilon)k_\star(d)
&\quad\Longrightarrow\quad
\Pr_{\rho\sim\pi_{d,k}}
[\rho\notin\operatorname{STAB}_n]\longrightarrow0,
\end{aligned}
\label{eq:magic-threshold-definition}
\end{equation}
as $d\rightarrow\infty$.

The connection between magic typicality and the geometry of the stabilizer polytope follows from the general theory of random induced states~\cite{AubrunSzarekYe2014}.
For the centered stabilizer polytope $K_n=\operatorname{STAB}_n-\mathbb{I}/d$, a state $\rho$ is stabilizer if and only if $\|\rho-\mathbb{I}/d\|_{K_n}\leq1$.
In the regime relevant to the transition, random induced states admit a Gaussian approximation,
\begin{equation}\label{eq:gaussian-induced}
\mathbb{E}
\left\|
\rho_{d,k}-\frac{\mathbb{I}}{d}
\right\|_{K_n}
=
(1+o(1))
\frac{\mathbb{E}\|G\|_{K_n}}{d\sqrt{k}},
\end{equation}
where $G$ is a standard Gaussian vector in the traceless Hermitian space.
The inradius bound in Theorem~\ref{thm:purity-radius} and the few-vertex volume estimate underlying Theorem~\ref{thm:volume-radius} provide, respectively, upper and lower bounds on the Gaussian gauge $\mathbb{E}\|G\|_{K_n}$, yielding
\begin{equation}
\Omega\!\left(\frac{d^2}{\log d}\right)
\leq
\mathbb{E}\|G\|_{K_n}
\leq
\mathcal{O}(d^2).
\label{eq:gaussian-width-bound}
\end{equation}
Together with the corresponding concentration estimate for random induced states~\cite{AubrunSzarekYe2014}, Eq.~\eqref{eq:gaussian-induced} shows that the transition occurs when this expectation crosses unity, leading to the threshold scale stated below, with detailed proof in Appendix~\ref{app:magic-typicality}:

\begin{theorem}[Magic typicality transition]
\label{thm:magic-typicality}
Let $d=2^n$. There exists a magic typicality threshold $k_\star(d)$ satisfying
\begin{equation}
\Omega\!\left(\frac{d^2}{\log^2d}\right)
\leq
k_\star(d)
\leq
\mathcal{O}(d^2).
\label{eq:magic-threshold-qubit}
\end{equation}
\end{theorem}

Theorem~\ref{thm:magic-typicality} shows that magic survives stronger environmental mixing than entanglement across a fixed balanced bipartition.
Consider an even number \(n\) of system qubits with total dimension \(d=2^n\), and fix a bipartition \(A:B\) with \(d_A=d_B=\sqrt d\).
The entanglement--separability transition across this cut occurs at an environment dimension of order \(d^{3/2}\), up to logarithmic factors~\cite{AubrunSzarekYe2014,aubrun2012phase}, whereas the magic threshold lies between \(\Omega(d^2/\log^2 d)\) and \(\mathcal O(d^2)\).
In terms of qubit counts, entanglement across $A$ and $B$ typically disappears when the environment reaches roughly \(3n/2\) qubits, whereas magic persists until the environment contains roughly \(2n\) qubits.
Hence, there is a broad regime between roughly $3n/2$ and $2n$ environment qubits in which random induced states are typically separable across $A:B$ while still possessing magic.

\section{Detectability of Magic}
\label{sec:detectability}
We now show that magic can become difficult to certify even while remaining typical.  In particular, commonly used criteria for mixed-state magic lose their detection power well before magic itself ceases to be typical. For example, the probability that the Triangle Criterion detects magic is already exponentially small when the environment dimension $k$ reaches a sufficiently large constant multiple of $n^2$~\cite{LiuHaugYe2026Triangle}.

One may also consider the criterion based on the normalized Pauli norm~\cite{campbell2011catalysis},
\begin{equation}
\mathcal D(\rho)
:=\frac{1}{d}\sum_{P\in\{\mathbb I,X,Y,Z\}^{\otimes n}}
\left|\operatorname{Tr}(P\rho)\right|,
\label{eq:pauli-norm}
\end{equation}
where $d=2^n$ and the sum includes the identity. Every pure stabilizer state has exactly $d$ nonzero Pauli expectation values, each of unit magnitude. By convexity, $\mathcal D(\rho)\leq 1$ for every $\rho\in\operatorname{STAB}_n$, so $\mathcal D(\rho)>1$ certifies magic. For random induced states, we show that this criterion undergoes a sharp detection transition at an environment dimension of order $d$, well below the scale at which magic itself disappears.

\begin{theorem}[Pauli norm detection transition]
\label{thm:pauli-detection}
Let $d=2^n$, for every $\epsilon\in(0,1)$,
\begin{equation}
\begin{aligned}
k\leq(1-\epsilon)\frac{2d}{\pi}
&\quad\Longrightarrow\quad
\Pr[\mathcal D(\rho)>1]\longrightarrow1,\\
k\geq(1+\epsilon)\frac{2d}{\pi}
&\quad\Longrightarrow\quad
\Pr[\mathcal D(\rho)>1]\longrightarrow0.
\end{aligned}
\label{eq:pauli-threshold}
\end{equation}
\end{theorem}

The proof, given in Appendix~\ref{app:pauli-detection}, is based on combining the exact Haar distribution of a Pauli expectation value with concentration of measure.
The result indicates that the detection threshold is   $\frac{2}{\pi}d$, whereas Theorem~\ref{thm:magic-typicality} places the disappearance of magic at a scale approaching $d^2$.
In particular, when $d\ll k\ll d^2/\log^2d$, random induced states are typically magical but cannot be detected by the Pauli norm.

\begin{figure}[t]
\centering
\includegraphics[width=0.46\textwidth]{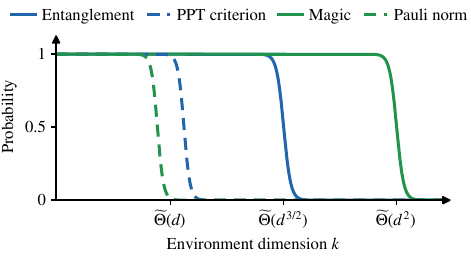}
\caption{Schematic comparison of resource existence and detection for random induced states \(\rho\sim\pi_{d,k}\), where \(d=2^n\) is the total system dimension and \(k\) is the environment dimension. Solid blue and green curves show the probabilities of entanglement and magic, respectively; dashed curves show detection by the PPT and Pauli norm criteria. Entanglement and partial transposition refer to a fixed balanced bipartition. }
\label{fig:thresholds}
\end{figure}

An analogy can be drawn with the separation between the positive partial transpose (PPT) criterion and entanglement itself.
For the same balanced bipartition, the probability of a negative partial transpose (NPT) drops from one to zero at $k\sim4d$~\cite{Aubrun2012PartialTransposition}, whereas entanglement persists to the $d^{3/2}$ scale up to logarithmic corrections~\cite{aubrun2012phase,AubrunSzarekYe2014}.
Figure~\ref{fig:thresholds} illustrates the comparison between these four transitions.
Although magic survives stronger environmental mixing than entanglement across this cut, persisting up to the scale $k=\widetilde{\Theta}(d^2)$, the Pauli norm criterion, like the PPT criterion, loses its detection power already at $k=\Theta(d)$.
This leaves a parametrically broader range of environment sizes in which magic remains typical yet is invisible to simple criteria such as the Pauli norm.

The ineffectiveness of these two specific magic criteria in this regime leaves open whether a sharper criterion could certify magic for a substantial fraction of random induced states at lower cost. To address this question, we consider protocols that use the expectation values of $M$ fixed single-copy observables~\cite{LiuHaugYe2026Triangle}. The observables are chosen independently of the unknown state, and their expectation values are assumed to be known exactly. A protocol may process these values arbitrarily, but it can certify magic only if no mixture of stabilizer states is compatible with all of them. The following theorem resolves the detection conjecture of Ref.~\cite{LiuHaugYe2026Triangle} by establishing a lower bound linear in the environment dimension.

\begin{theorem}
[Difficulty of single-copy magic detection]
\label{coro:magic-detection}
Let $\rho\sim\pi_{d,k}$. Any single-copy linear detection protocol based on fixed observables that certifies the magic of $\rho$ with constant probability requires
\begin{equation}
M=\Omega(k)
\end{equation}
observables.
\end{theorem}

The proof, given in Appendix~\ref{app:magic-detection}, relies on the inradius bound of Theorem~\ref{thm:purity-radius}. Suppose that a state inside the stabilizer ball around the maximally mixed state reproduces all $M$ measured expectation values of the unknown state. Since every state in this ball is a mixture of stabilizer states, the measured data cannot certify magic. We show that when $k\gg M$, such a matching state exists with high probability for a random induced state. Certifying magic with constant probability therefore requires $M=\Omega(k)$ distinct observables.

The gap between typicality and detectability is already evident when the environment is comparable in size to the system. For example, over a broad range of environment sizes from roughly $n$ to nearly $2n$ qubits, the reduced state is magic with high probability, yet the Pauli-norm and Triangle criteria considered above typically fail to detect it. 
Throughout this same regime, any protocol based on fixed single-copy observables requires exponentially many distinct observables to certify magic with constant probability, even when their expectation values are known exactly.

\section{Facet Complexity of Stabilizer Polytopes}
\label{sec:facets}

As discussed above, the $n$-qubit stabilizer states constitute the vertices of a full-dimensional polytope in the $(d^2-1)$-dimensional real affine space of density operators.
Given the dimension and number of vertices, it is important to
determine how many facets the stabilizer polytope can have as this quantifies the complexity of characterizing it by linear
inequalities.
For general polytopes, this question is governed by McMullen's upper-bound theorem~\cite{mcmullen1970maximum}.
Since the number of pure $n$-qubit stabilizer states satisfies $\log(N_d)=\Theta(\log^2d)$ and the dimension of the stabilizer polytope is $d^2-1$, this general result implies
\begin{equation}
F_d
\leq
\exp\!\left[\mathcal{O}\!\left(d^2\log^2d \right)\right],
\label{eq:facet-general-upper}
\end{equation}
where $F_d$ denotes the number of facets of $\operatorname{STAB}_n$~\cite{ZurelOkayRaussendorf2024Bits}.
In contrast, the best previously known lower bound, obtained from contextuality arguments, only gives $\log_2 F_d=\Omega(\log^2d)$~\cite{Karanjai2018Contextuality,ZurelOkayRaussendorf2024Bits}.
Hence, a huge gap remains between the known lower bound and the combinatorially allowed number of facets.

The geometric results developed above allow us to substantially narrow this gap.
The key idea is again most transparent in the polar geometry, where facets of the centered stabilizer polytope are in one-to-one correspondence with vertices of its polar.
Theorem~\ref{thm:purity-radius} implies that the polar is contained in a Euclidean ball of radius $\mathcal{O}(d)$.
On the other hand, the reverse Santal\'o inequality of Bourgain and Milman~\cite{bourgain1987new} indicates that the product of the
volume radii of any centrally symmetric convex body and its polar is bounded below by a universal positive constant. 
After a standard symmetrization, combining this inequality with the
volume-radius bound of Theorem~\ref{thm:volume-radius} implies that
the polar stabilizer polytope must have a parametrically large volume
radius despite being confined inside this Euclidean ball.  
A high-dimensional polytope that occupies such a large volume within a bounded ball cannot have only a small number of vertices.
Since these vertices correspond precisely to the facets of the original stabilizer polytope, this immediately leads to a strong lower bound on its facet complexity, with detailed proof left in Appendix~\ref{app:facet-complexity}.

\begin{theorem}[Facet complexity]
\label{thm:facet-complexity}
Let $d=2^n$. The number $F_d$ of facets of the $n$-qubit stabilizer polytope satisfies
\begin{equation}
\exp\!\left[
\Omega\!\left(
\frac{d^2}{\log^2 d}
\right)
\right]
\leq
F_d
\leq
\exp\!\left[
\mathcal{O}\!\left(
d^2\log^2 d
\right)
\right].
\label{eq:facet-bound-d}
\end{equation}
\end{theorem}

Theorem~\ref{thm:facet-complexity} shows that the number $F_d$ of facets of the $n$-qubit stabilizer polytope grows doubly exponentially in $n$. On a logarithmic scale, its boundary complexity is close to the highest possible for a polytope with the same dimension and number of vertices: $\log F_d$ is within a polylogarithmic factor in $d$ of its maximal possible value. Even after identifying facets related by Clifford conjugation, the number of inequivalent facets remains doubly exponential, since the $n$-qubit Clifford group modulo global phases has only $2^{\Theta(n^2)}$ elements. 
These results show that enumerating all facets of the stabilizer polytope is an impractical route to magic detection.
They motivate approaches that exploit the structure and symmetries of stabilizer states to reduce the computational cost.
For example, given a classical description of a density matrix, one can compute its robustness of magic in $2^{\mathcal O(n^2)}$ time~\cite{leone2026unbearable}.
For structured families, such as tensor powers of standard single-qubit magic states, exploiting symmetries can substantially reduce the computational cost~\cite{Heinrich2019robustnessofmagic}.

Our lower bound substantially strengthens the existing estimate $\log F_d=\Omega(\log^2 d)$ obtained from contextuality-based bounds~\cite{Karanjai2018Contextuality,ZurelOkayRaussendorf2024Bits}. It also clarifies how much of the boundary can be described by closed and noncontextual (CNC) sets of Pauli observables. Such sets are closed under products of commuting observables and admit consistent assignments of measurement outcomes. For a single qudit of odd prime dimension, every facet admits a CNC description~\cite{howard2014contextuality}. For $n$ qubits, however, the CNC facets form exactly $n$ Clifford orbits~\cite{zurel2026basis}, thus account for at most $2^{\mathcal O(n^2)}$ facets, a vanishing fraction of the total. Therefore, even up to Clifford equivalence, almost all multiqubit facet witnesses lie outside the CNC family.

\section{Discussion}
\label{sec:discussion}

Our results lead to two main take-home messages. On the practical side, we establish a separation between the typicality and detectability of magic under environmental noise. Over a broad range of environment sizes, random induced states are magic with high probability, yet commonly used criteria, including the Triangle and Pauli norm criteria, typically fail to detect it. Even with exact expectation values, any protocol based on fixed single-copy observables requires a number of distinct observables exponential in the number of system qubits to certify magic with constant probability in this regime.
On the geometric side, as a high-dimensional polytope, the stabilizer polytope exhibits near-extremal geometric complexity allowed by its dimension and number of vertices. Its volume radius is within a logarithmic factor of the maximal scale permitted for a polytope with the same dimension and vertex count, while the logarithm of its number of facets is determined up to polylogarithmic factors. Taken together, these results establish geometrically that, despite their highly structured and discrete nature, pure stabilizer states form a remarkably well-spread, near-extremal configuration in operator space.

A natural open problem is to determine the exact inradius of the stabilizer polytope.
The Triangle Criterion gives a boundary construction at purity $1/(d-1/2)$, whereas our result guarantees stabilizerness up to $1/(d-a_\ast)$ with $a_\ast=0.458327\cdots$.
Closing this remaining constant gap would completely determine the
purity threshold for the onset of magic.
Such an improvement will likely require exploiting the discrete stabilizer structure beyond the spectral information captured by the present second- and third-moment analysis.

Another important direction is to understand how the geometric transition identified here manifests itself in concrete physical systems.
Random induced states provide a natural universal ensemble, but experimentally and physically relevant mixed states are typically generated by local Hamiltonian dynamics, coupling to an environment, or thermalization.
It would therefore be interesting to investigate whether analogous magic transitions occur in reduced states of chaotic many-body systems, noisy quantum circuits, open system dynamics, or finite-temperature Gibbs states.
In particular, one may ask whether there exist broad physical regimes in which entanglement has already disappeared while magic remains present~\cite{wei2026entirely}, as suggested by the separation between the typical entanglement and magic thresholds found here. Whether the surviving magic can be distilled or exploited for computation is a further question.

\begin{acknowledgments}
We thank Hao Dai, Zhenyu Du, Shunlong Luo, and Fuchuan Wei for insightful discussions.
Z.-W.L.~is supported in part by NSFC under Grant No.~12475023, Dushi Program, and startup funding from YMSC.
\end{acknowledgments}


%


\appendix

\section*{APPENDICES}

We will provide all proofs of the results stated in the main text.
Throughout the appendices, $d=2^n$ denotes the Hilbert space dimension and $m=d^2-1$ the real dimension of the affine state space.
We retain $K_n=\operatorname{STAB}_n-\mathbb{I}/d$ and $a_\ast=0.458327\cdots$ from the main text.
The number of pure stabilizer states will be denoted by $N_d:=d\prod_{j=1}^{\log_2 d}(2^j+1)$, for which $\log N_d=\Theta(\log^2 d)$.
Whenever a centrally symmetric body is needed, we use $\widetilde K_n:=\operatorname{conv}(K_n\cup-K_n)$.

\section{Proof of Theorem~\ref{thm:purity-radius}}
\label{app:purity-radius}

Let $\mathcal{S}_n$ denote the set of pure $n$-qubit stabilizer-state projectors and define the trace-one stabilizer-positive polytope
\begin{equation}
\Lambda_n
=
\left\{
A=A^\dagger:
\operatorname{Tr}A=1,\;
\operatorname{Tr}(A\sigma)\geq0
\ \text{for all}\ 
\sigma\in\mathcal{S}_n
\right\}.
\end{equation}
We write $Q_d:=\max_{A\in\Lambda_n}\operatorname{Tr}(A^2)$.

\paragraph{Dual formulation.}
Let $c=\mathbb{I}/d$.
The polar of the centered stabilizer polytope satisfies
$K_n^\circ=-d(\Lambda_n-c)$.
Indeed, for any trace-one Hermitian operator $A$ and any pure stabilizer state $\sigma$, one has
$\operatorname{Tr}(A\sigma)=1/d+\langle A-c,\sigma-c\rangle$.
Hence $-d(A-c)\in K_n^\circ$ if and only if
$\operatorname{Tr}(A\sigma)\geq0$ for every $\sigma\in\mathcal{S}_n$.

We first note that the Hilbert--Schmidt inradius of $K_n$ is attained by a ball centered at the origin.
Indeed, if $Z+rB_2^m\subseteq K_n$, Pauli invariance implies
$PZP^\dagger+rB_2^m\subseteq K_n$ for every Pauli operator $P$.
Averaging these inclusions and using convexity of $K_n$, together with the Pauli-twirling identity
$\frac{1}{|\mathcal P_n|}\sum_{P\in\mathcal P_n}PZP^\dagger=0$
for traceless $Z$, gives $rB_2^m\subseteq K_n$.
Thus allowing an arbitrary center cannot increase the radius of the largest inscribed ball.

The inradius can therefore be expressed directly through the polar as
\begin{equation}
r_{\mathrm{in}}(K_n)
=
\frac{1}{\max_{X\in K_n^\circ}\|X\|_2}
=
\frac{1}{\sqrt{d(dQ_d-1)}}.
\label{eq:rin-Qd}
\end{equation}
Thus it remains to upper-bound $Q_d$ by a dimension-independent constant.

\paragraph{Pauli-compression recursion.}
Let $A\in\Lambda_n$ and write $q=\operatorname{Tr}(A^2)$, $t=\operatorname{Tr}(A^3)$, and $M=Q_{d/2}$.
For every nonidentity Hermitian Pauli operator $P$, let $\Pi_\pm=(\mathbb{I}\pm P)/2$.
Choose Clifford isometries $V_\pm:\mathbb{C}^{d/2}\rightarrow\operatorname{Ran}(\Pi_\pm)$ satisfying $V_\pm V_\pm^\dagger=\Pi_\pm$, and define $B_\pm=V_\pm^\dagger A V_\pm$.
Every stabilizer state of the compressed $(n-1)$-qubit system is mapped by $V_\pm$ to an $n$-qubit stabilizer state.
Moreover, $\operatorname{Tr}B_\pm=(1\pm\operatorname{Tr}(PA))/2\geq0$, since $(\mathbb{I}\pm P)/d\in\operatorname{STAB}_n$.
{If $\operatorname{Tr}B_\pm>0$, the definition of $M$ applies to $B_\pm/\operatorname{Tr}B_\pm$. If $\operatorname{Tr}B_\pm=0$, then, for each fixed $B_\pm$, every $R\in\mathcal P_{n-1}\setminus\{\mathbb I\}$, and both $\eta\in\{\pm1\}$, stabilizer positivity gives $0\leq\operatorname{Tr}[B_\pm(\mathbb I+\eta R)/(d/2)]=\eta\operatorname{Tr}(B_\pm R)/(d/2)$. Hence all Pauli coefficients of $B_\pm$ vanish, so $B_\pm=0$. Therefore, in both cases,
\begin{equation}
\operatorname{Tr}(B_\pm^2)
\leq
M[\operatorname{Tr}(B_\pm)]^2.
\label{eq:homogeneous-bound}
\end{equation}}

Set $x_P=\operatorname{Tr}(PA)$, $y_P=\operatorname{Tr}(PA^2)$, and $c_P=\operatorname{Tr}(PAPA)$.
Then $|x_P|\leq1$ and $\operatorname{Tr}B_\pm=(1\pm x_P)/2$, $\operatorname{Tr}(B_\pm^2)=(q+c_P\pm2y_P)/4$.
Applying Eq.~\eqref{eq:homogeneous-bound} to both signs yields
\begin{equation}
q+c_P+2|y_P-Mx_P|
\leq
M(1+x_P^2).
\label{eq:single-pauli-compression}
\end{equation}

Summing Eq.~\eqref{eq:single-pauli-compression} over all nonidentity Pauli operators and using Pauli orthogonality and the Pauli twirl gives $\sum_{P\neq\mathbb{I}}x_P^2=dq-1$ and $\sum_{P\neq\mathbb{I}}c_P=d-q$.
Furthermore, using $|x_P|\leq1$ and $\sum_{P\neq\mathbb{I}}x_Py_P=d\,t-q$,
\begin{equation}
\sum_{P\neq\mathbb{I}}|y_P-Mx_P|
\geq
-\sum_{P\neq\mathbb{I}}x_P(y_P-Mx_P)
=
(dM+1)q-M-dt.
\end{equation}
Substitution into the summed compression inequalities gives
\begin{equation}
d(q-M)
\leq
2t-Mq-1.
\label{eq:moment-recursion}
\end{equation}
We also note that $q=(1+\sum_{P\neq\mathbb{I}}x_P^2)/d\leq d$.

\paragraph{A sharp spectral bound on the third moment.}
For every Hermitian $A$ with $\operatorname{Tr}A=1$ and $\operatorname{Tr}(A^2)=q$,
\begin{equation}
\operatorname{Tr}(A^3)
\leq
T_d(q)
:=
\frac{3q}{d}
-\frac{2}{d^2}
+
\frac{d-2}{\sqrt{d(d-1)}}
\left(q-\frac1d\right)^{3/2}.
\label{eq:Td}
\end{equation}
To see this, write the eigenvalues as $\lambda_i=1/d+z_i$, so that $\sum_i z_i=0$ and $\sum_i z_i^2=q-1/d$.
The Lagrange equations imply that an extremizer of $\sum_i z_i^3$ has at most two distinct values.
If the larger value occurs $k$ times, then $\sum_i z_i^3=(q-1/d)^{3/2}(d-2k)/\sqrt{dk(d-k)}$, which is maximized at $k=1$ and yields Eq.~\eqref{eq:Td}.

\paragraph{Scalar recursion.}
Suppose $Q_{d/2}\leq U$ and set $q=Q_d$.
Define $H_{d,U}(x):=d(x-U)+Ux+1-2T_d(x)$.
Equation~\eqref{eq:moment-recursion} implies $H_{d,U}(q)\leq0$.
Replacing $M=Q_{d/2}$ by $U\geq M$ preserves the inequality because the change is $(U-M)(q-d)\leq0$.
For $d\geq8$, $H_{d,U}$ is strictly increasing on $U\leq x\leq d$; indeed, $H_{d,U}'(x)=d+U-6/d-3(d-2)\sqrt{x-1/d}/\sqrt{d(d-1)}>0$ on this interval.
Therefore, any $V\geq U$ satisfying $H_{d,U}(V)\geq0$ certifies $Q_d\leq V$.

It is obvious that $Q_2=2$.
The exact two-qubit value is $Q_4=2$ {according to the exhaustive
enumeration of the two-qubit stabilizer polytope facets in
Ref.~\cite{reichardt2006quantum}}.
Iterating the preceding criterion gives the representative bounds
\begin{equation}
\begin{array}{c|c}
d & \text{upper bound on }Q_d{\text{ (upward rounded)}} \\
\hline
8 & 67/32=2.093750\\
16 & 2.138430834\\
32 & 2.160289062\\
64 & 2.171104560\\
128 & 2.176484702\\
256 & 2.179167967\\
4096 & 2.181679439\\
262144 & 2.181844116\\
2^{24} & 2.181846689
\end{array}
\label{eq:recursive-table}
\end{equation}
All intermediate steps up to $d=2^{24}$ are verified using exact rational arithmetic.
For rational $U$ and $V$, the only irrational term in $H_{d,U}(V)$ is proportional to $(V-1/d)^{3/2}$, so $H_{d,U}(V)\geq0$ can be certified by a single rational squaring inequality.

\paragraph{Control of the infinite tail.}
Using $|\operatorname{Tr}(A^3)|\leq q^{3/2}$ in Eq.~\eqref{eq:moment-recursion}, define
\begin{equation}
J_{d,U}(x):=d(x-U)+Ux+1-2x^{3/2}.
\end{equation}
Again $J_{d,U}(Q_d)\leq0$ whenever $Q_{d/2}\leq U$.
Let $y>1$ solve $y^3-y^2-y-1=0$ and set $R=y^2=3.3829757679\cdots$.
Then
\begin{equation}
J_{d,R}(R)
=
R^2+1-2R^{3/2}
=
(y-1)(y^3-y^2-y-1)
=
0.
\end{equation}
Since
$J'_{d,U}(x)=d+U-3\sqrt{x}\geq d+2-3\sqrt d>0$
for $d\geq8$ and $x\leq d$,
$J_{d,U}$ is strictly increasing on the relevant interval.
Therefore, induction from $Q_4=2<R$ gives $Q_d\leq R$.

Now set $B=a_*^{-1}=2.181847\cdots$ and $C=1.081952$, which satisfy
$C+2B+1-2B^{3/2}>0$.
Suppose $Q_{d/2}\leq U$ and define $V=U+C/d$.
Since $U\geq2$ and $2x+1-2x^{3/2}$ is decreasing for $x\geq2$, whenever $V\leq B$,
\begin{equation}
J_{d,U}(V)
\geq
C+2V+1-2V^{3/2}
\geq
C+2B+1-2B^{3/2}
>0.
\end{equation}
Hence $Q_d<V$.
At the end of the finite certificate, $D=2^{24}$ and the certified bound $U_D$ satisfies $B-U_D>C/D$.
The total possible increase over all subsequent dimension doublings is
$C/(2D)+C/(4D)+\cdots=C/D$.
Therefore
\begin{equation}
Q_d<a_*^{-1},
\quad
\text{for every qubit dimension }d.
\label{eq:Q-final}
\end{equation}

\paragraph{Conversion to a purity bound.}
Combining Eqs.~\eqref{eq:rin-Qd} and~\eqref{eq:Q-final} gives
\begin{equation}
r_{\mathrm{in}}(K_n)
>
\frac{1}{\sqrt{d(a_*^{-1}d-1)}}
=
\sqrt{\frac{a_*}{d(d-a_*)}}.
\end{equation}
Finally, $\|\rho-\mathbb{I}/d\|_2^2=\operatorname{Tr}(\rho^2)-1/d$.
Hence every state satisfying
$\operatorname{Tr}(\rho^2)\leq1/(d-a_*)$
belongs to $\operatorname{STAB}_n$, proving the claimed purity threshold and lower inradius bound in Theorem~\ref{thm:purity-radius}.

\section{Proof of Theorem~\ref{thm:volume-radius}}
\label{app:volume-radius}

By Theorem~\ref{thm:purity-radius}, $r_0B_2^m\subseteq K_n$ with $r_0=\sqrt{a_\ast/[d(d-a_\ast)]}$, and therefore $\operatorname{vrad}(K_n)\geq r_0$.
For the upper bound, we use the classical few-vertex estimate~\cite{barany1988approximation,carl1988gelfand}: if an $m$-dimensional polytope $P\subseteq B_2^m$ has $N\geq m$ vertices, then
\begin{equation}
\operatorname{vrad}(P)
\leq
C_0\sqrt{\frac{\log(eN/m)}{m}},
\label{eq:few-vertex}
\end{equation}
where $C_0>0$ is universal.
Each vertex $v_\phi=\ketbra{\phi}{\phi}-\mathbb{I}/d$ of $K_n$ satisfies $\|v_\phi\|_2^2=1-1/d<1$, so $K_n\subset B_2^m$.
Using $\log N_d=\Theta(\log^2 d)$ and $m=\Theta(d^2)$ in Eq.~\eqref{eq:few-vertex} gives $\operatorname{vrad}(K_n)\leq C\log d/d$.
Thus
\begin{equation}
\sqrt{\frac{a_\ast}{d(d-a_\ast)}}
\leq
\operatorname{vrad}(K_n)
\leq
C\frac{\log d}{d}.
\end{equation}
This proves Theorem~\ref{thm:volume-radius}.

\section{Proof of Theorem~\ref{thm:magic-typicality}}
\label{app:magic-typicality}

Let $\|\cdot\|_{K_n}$ denote the gauge of the centered stabilizer polytope $K_n$.
For a convex body $L$, we denote its support function by
$h_L(X):=\sup_{Y\in L}\langle X,Y\rangle$.
Recall the gauge--support duality $\|X\|_L=h_{L^\circ}(X)$.
Let $G$ be a standard Gaussian vector in the $m=d^2-1$ dimensional real vector space of traceless Hermitian operators equipped with the Hilbert--Schmidt inner product, and define $\gamma_m:=\mathbb{E}\|G\|_2$.
We define the transition scale
\begin{equation}
k_\star(d)
=
\left(
\frac{\mathbb{E}\|G\|_{K_n}}{d}
\right)^2.
\label{eq:kstar-proof}
\end{equation}

We first explain why this quantity determines the transition scale.
A Gaussian approximation theorem for random induced states~\cite{AubrunSzarekYe2014} states that, in the joint asymptotic regime $d\to\infty$ and $k/d\to\infty$,
\begin{equation}
\mathbb{E}
\left\|
\rho_{d,k}-\frac{\mathbb{I}}{d}
\right\|_{K_n}
=
(1+o(1))
\frac{\mathbb{E}\|G\|_{K_n}}{d\sqrt{k}}
=
(1+o(1))
\sqrt{\frac{k_\star(d)}{k}}.
\label{eq:gauge-expectation}
\end{equation}
Thus, provided $k_\star(d)/d\to\infty$, the Gaussian approximation is valid in the vicinity of $k_\star(d)$.
Since $\rho_{d,k}\in\operatorname{STAB}_n$ if and only if
$\|\rho_{d,k}-\mathbb{I}/d\|_{K_n}\leq1$, Eq.~\eqref{eq:gauge-expectation}, together with the corresponding concentration estimate~\cite{AubrunSzarekYe2014}, implies a sharp transition when $k$ crosses $k_\star(d)$.
We will verify below that $k_\star(d)/d\to\infty$ and determine its scaling up to logarithmic factors.

\paragraph{Upper bound.}

Theorem~\ref{thm:purity-radius} gives
$r_{\mathrm{in}}(K_n)B_2^m\subseteq K_n$, where
$r_{\mathrm{in}}(K_n)\geq\sqrt{a_\ast/[d(d-a_\ast)]}$.
Taking polars reverses the inclusion and therefore gives
$K_n^\circ\subseteq r_{\mathrm{in}}(K_n)^{-1}B_2^m$.
Using the gauge--support duality and the Cauchy--Schwarz inequality,
\begin{equation}
\|G\|_{K_n}
=
h_{K_n^\circ}(G)
=
\sup_{Y\in K_n^\circ}\langle G,Y\rangle
\leq
\frac{\|G\|_2}{r_{\mathrm{in}}(K_n)}.
\label{eq:gauge-inradius}
\end{equation}
Since $G$ is a standard Gaussian vector in $m$ real dimensions, $\mathbb{E}\|G\|_2^2=m$.
Hence Jensen's inequality gives $\gamma_m\leq\sqrt{m}<d$.
Taking expectations in Eq.~\eqref{eq:gauge-inradius} and substituting into Eq.~\eqref{eq:kstar-proof}, we obtain
\begin{equation}
k_\star(d)
\leq
\frac{\gamma_m^2}
{d^2r_{\mathrm{in}}(K_n)^2}
\leq
\frac{d(d-a_\ast)}{a_\ast}
=
\mathcal{O}(d^2).
\label{eq:kstar-upper}
\end{equation}

\paragraph{Lower bound.}

We now use the global geometry of the stabilizer polytope.
Recall the symmetrized body
$\widetilde K_n=\operatorname{conv}(K_n\cup-K_n)$.
It has at most $2N_d$ vertices, all contained in the Euclidean unit ball.
The few-vertex estimate in Eq.~\eqref{eq:few-vertex}, together with $\log N_d=\Theta(\log^2 d)$ and $m=\Theta(d^2)$, gives
\begin{equation}
\operatorname{vrad}(\widetilde K_n)
=
\mathcal{O}\!\left(
\frac{\log d}{d}
\right).
\label{eq:sym-vrad-upper}
\end{equation}

To pass from the volume of $\widetilde K_n$ to that of its polar, we use the Bourgain--Milman reverse Santal\'o inequality~\cite{bourgain1987new}.
For any centrally symmetric convex body $L\subset\mathbb{R}^m$ containing the origin in its interior,
\begin{equation}
\operatorname{vrad}(L)
\operatorname{vrad}(L^\circ)
\geq
c_{\mathrm{BM}},
\label{eq:reverse-santalo}
\end{equation}
where $c_{\mathrm{BM}}>0$ is a universal constant.
Since $\widetilde K_n$ is centrally symmetric, Eqs.~\eqref{eq:sym-vrad-upper} and~\eqref{eq:reverse-santalo} imply
\begin{equation}
\operatorname{vrad}(\widetilde K_n^\circ)
=
\Omega\!\left(
\frac{d}{\log d}
\right).
\label{eq:sym-polar-vrad-typicality}
\end{equation}

We next convert this volume-radius bound into a Gaussian-width bound using Urysohn's inequality~\cite{MilmanSchechtman1986}.
For any convex body $L\subset\mathbb{R}^m$,
\begin{equation}
\operatorname{vrad}(L)
\leq
\frac{\mathbb{E}h_L(G)}{\gamma_m}.
\label{eq:urysohn-gaussian}
\end{equation}
Equivalently, $\mathbb{E}h_L(G)\geq\gamma_m\operatorname{vrad}(L)$.
Applying Eq.~\eqref{eq:urysohn-gaussian} to $L=\widetilde K_n^\circ$ gives
\begin{equation}
\mathbb{E}
h_{\widetilde K_n^\circ}(G)
\geq
\gamma_m
\operatorname{vrad}(\widetilde K_n^\circ).
\label{eq:urysohn-application}
\end{equation}

For a standard Gaussian vector in $m$ dimensions,
$\gamma_m=\sqrt{2}\,\Gamma[(m+1)/2]/\Gamma(m/2)=\Theta(\sqrt{m})$.
Since $m=d^2-1$, we therefore have $\gamma_m=\Theta(d)$.
Combining this with Eq.~\eqref{eq:sym-polar-vrad-typicality}, and using the gauge--support duality,
\begin{equation}
\mathbb{E}\|G\|_{\widetilde K_n}
=
\mathbb{E}h_{\widetilde K_n^\circ}(G)
=
\Omega\!\left(
\frac{d^2}{\log d}
\right).
\label{eq:sym-gaussian-width}
\end{equation}

Finally, since $K_n\subseteq\widetilde K_n$, monotonicity of the gauge under set inclusion gives
$\|G\|_{K_n}\geq\|G\|_{\widetilde K_n}$.
Hence
\begin{equation}
\mathbb{E}\|G\|_{K_n}
=
\Omega\!\left(
\frac{d^2}{\log d}
\right),
\qquad
k_\star(d)
=
\Omega\!\left(
\frac{d^2}{\log^2 d}
\right).
\label{eq:kstar-lower}
\end{equation}

Combining Eqs.~\eqref{eq:kstar-upper} and~\eqref{eq:kstar-lower}, we obtain
\begin{equation}
\Omega\!\left(
\frac{d^2}{\log^2 d}
\right)
\leq
k_\star(d)
\leq
\mathcal{O}(d^2).
\label{eq:kstar-final}
\end{equation}
In particular,
$k_\star(d)/d=\Omega(d/\log^2 d)\to\infty$.
Therefore the condition $k/d\to\infty$ required by the Gaussian approximation in Eq.~\eqref{eq:gauge-expectation} is automatically satisfied throughout the transition window $k=\Theta(k_\star(d))$.
Together with the concentration estimate discussed above, this establishes the sharp magic transition stated in Theorem~\ref{thm:magic-typicality}.

It remains to extend the lower-side conclusion to all $k\leq(1-\varepsilon)k_\star(d)$. 
We divide this range into three regimes.
First, if $k<d$, then $\rho_{d,k}$ has rank $k$ almost surely and its support is Haar distributed on the Grassmannian of $k$-dimensional subspaces.
On the other hand, the support of any rank-deficient state in $\operatorname{STAB}_n$ must be spanned by a subset of the finitely many pure stabilizer states.
There are therefore only finitely many possible such supports, and a Haar-random $k$-dimensional subspace coincides with any of them with probability zero.
Hence
$\Pr_{\rho\sim\pi_{d,k}}[\rho\in\operatorname{STAB}_n]=0$
for $k<d$.
For the intermediate regime $d\leq k<d\log d$, we use the density-comparison estimate for induced measures~\cite{AubrunSzarekYe2014}, which gives, for any measurable set $\mathcal K$ in the state space,
\begin{equation}
\Pr_{\rho\sim\pi_{d,k}}[\rho\in\mathcal K]^{1/m}
\leq
C\sqrt{\frac{k}{d}}\,
\Pr_{\rho\sim\pi_{d,d}}[\rho\in\mathcal K]^{1/m},
\end{equation}
where $m=d^2-1$.
Since $\pi_{d,d}$ is the Hilbert--Schmidt uniform measure, Theorem~\ref{thm:volume-radius} and
$\operatorname{vrad}(\mathcal D_d-\mathbb I/d)=\Theta(d^{-1/2})$, where $\mathcal D_d$ represents the set of density matrices,
imply
$\Pr_{\pi_{d,d}}[\rho\in\operatorname{STAB}_n]^{1/m}
=O(\log d/\sqrt d)$.
Consequently, throughout $d\leq k<d\log d$,
\begin{equation}
\Pr_{\rho\sim\pi_{d,k}}[\rho\in\operatorname{STAB}_n]^{1/m}
=
O\!\left(\frac{\log^{3/2}d}{\sqrt d}\right)
\longrightarrow0,
\end{equation}
and therefore the stabilizer probability itself vanishes asymptotically.

Finally, consider $d\log d\leq k\leq(1-\varepsilon)k_\star(d)$.
Here $k/d\geq\log d\to\infty$, so the Gaussian approximation in Eq.~\eqref{eq:gauge-expectation} and the corresponding concentration estimate apply throughout this regime.
Moreover,
\begin{equation}
\mathbb E
\left\|
\rho_{d,k}-\frac{\mathbb I}{d}
\right\|_{K_n}
=
(1+o(1))
\sqrt{\frac{k_\star(d)}{k}}
\geq
\frac{1+o(1)}{\sqrt{1-\varepsilon}}
>1,
\end{equation}
so the gauge is bounded away from the stabilizer boundary with overwhelming probability.
It follows that
$\Pr_{\rho\sim\pi_{d,k}}[\rho\notin\operatorname{STAB}_n]\to1$.
Together with the two smaller-$k$ regimes above, this establishes the lower side of the threshold for every
$k\leq(1-\varepsilon)k_\star(d)$ and completes the proof of Theorem~\ref{thm:magic-typicality}.

\section{Proof of Theorem~\ref{thm:pauli-detection}}
\label{app:pauli-detection}

Let $\rho\sim\pi_{d,k}$, where $d=2^n$. To sample $\rho$ from this ensemble, draw a Haar-random pure state $\ket{\Psi}$ on $\mathbb C^d\otimes\mathbb C^k$ and trace out the $k$-dimensional environment, so that $\rho=\operatorname{Tr}_{\mathbb C^k}\!\left(\ketbra{\Psi}{\Psi}\right)$.
We first compute the mean Pauli norm and then show that it concentrates around this mean.

For any nonidentity Pauli operator $P$, the observable $P\otimes\mathbb I_k$ has eigenvalues $\pm1$, each with multiplicity $dk/2$. Let $B$ be the squared norm of the projection of $\ket{\Psi}$ onto its $+1$ eigenspace. By the same Haar calculation as in Ref.~\cite{TurkeshiDymarskySierant2025Pauli}, $B\sim\operatorname{Beta}(dk/2,dk/2)$, and hence
\[
x_P:=\operatorname{Tr}(P\rho)
=\langle\Psi|P\otimes\mathbb I_k|\Psi\rangle
=2B-1.
\]
Its first absolute moment is
\begin{equation}
\mathbb E|x_P|
=\frac{\Gamma((dk+1)/2)}{\sqrt{\pi}\,\Gamma(dk/2+1)},
\end{equation}
where $\Gamma$ denotes the gamma function. By linearity of expectation, summing over the $d^2-1$ nonidentity Paulis and including the identity contribution gives
\begin{equation}
\mu_{d,k}:=\mathbb E\mathcal D(\rho)
=\frac{1}{d}+\frac{d^2-1}{d}
\frac{\Gamma((dk+1)/2)}{\sqrt{\pi}\,\Gamma(dk/2+1)}.
\label{eq:pauli-mean}
\end{equation}
This calculation does not require the Pauli coefficients to be independent. Stirling's formula then yields, uniformly for $k\geq1$ as $d\to\infty$,
\begin{equation}
\mu_{d,k}
=\frac{1}{d}
+\left(1-\frac{1}{d^2}\right)
\sqrt{\frac{2d}{\pi k}}
\left[1+\mathcal O\!\left(\frac{1}{dk}\right)\right].
\label{eq:pauli-mean-asymptotic}
\end{equation}

For two density operators $\rho$ and $\sigma$, the Cauchy--Schwarz inequality and Pauli orthogonality give
\begin{equation}
\begin{aligned}
|\mathcal D(\rho)-\mathcal D(\sigma)|
&\leq\frac1d\sum_P|\operatorname{Tr}[P(\rho-\sigma)]|\\
&\leq\sqrt{d}\,\|\rho-\sigma\|_2.
\end{aligned}
\end{equation}
For reductions of unit vectors $\ket\Psi$ and $\ket\Phi$, trace-norm contractivity under partial trace further gives
\begin{equation}
\|\rho_\Psi-\rho_\Phi\|_2
\leq\|\rho_\Psi-\rho_\Phi\|_1
\leq2\|\ket\Psi-\ket\Phi\|_2.
\end{equation}
Thus $F(\Psi):=\mathcal D(\rho_\Psi)$ is $2\sqrt d$-Lipschitz on the real unit sphere $S^{2dk-1}$. Applying the mean-centered form of L\'evy's lemma~\cite{Levy1922} gives
\begin{equation}
\Pr\!\left[\left|\mathcal D(\rho)-\mu_{d,k}\right|>t\right]
\leq C\exp(-ckt^2),\qquad t>0,
\label{eq:pauli-concentration}
\end{equation}
where $C,c>0$ are universal constants. The factor $k$ in the exponent comes from dividing the real ambient dimension $2dk$ by the squared Lipschitz bound $(2\sqrt d)^2$.

Suppose first that \(k/d\to\alpha\in(0,\infty)\). By
Eq.~\eqref{eq:pauli-mean-asymptotic}, the mean \(\mu_{d,k}\)
converges to \(\ell_\alpha:=\sqrt{2/(\pi\alpha)}\).
For any fixed \(\delta>0\), we therefore have
\(|\mu_{d,k}-\ell_\alpha|<\delta/2\) for all sufficiently large \(d\).
The triangle inequality and Eq.~\eqref{eq:pauli-concentration}
then give
\begin{equation}
\begin{aligned}
\Pr\!\left[|\mathcal D(\rho)-\ell_\alpha|>\delta\right]
&\leq \Pr\!\left[|\mathcal D(\rho)-\mu_{d,k}|>\delta/2\right] \\
&\leq C\exp(-ck\delta^2/4)\longrightarrow 0,
\end{aligned}
\label{eq:pauli-limit-tail}
\end{equation}
since \(k\to\infty\). Thus \(\mathcal D(\rho)\) converges in
probability to \(\ell_\alpha\). This limit exceeds the detection
boundary \(1\) for \(\alpha<2/\pi\) and falls below it for
\(\alpha>2/\pi\).

To bound the detection probability uniformly on either side of
the threshold, fix \(0<\epsilon<1\). Rewriting
Eq.~\eqref{eq:pauli-mean-asymptotic} gives
\begin{equation}
\sqrt{\frac{k}{d}}\,(\mu_{d,k}-1)=\sqrt{\frac{2}{\pi}}\bigl[1+\mathcal O(d^{-1})\bigr] -\left(1-\frac{1}{d}\right)\sqrt{\frac{k}{d}},
\label{eq:pauli-scaled-mean-gap}
\end{equation}
uniformly for \(k\geq1\). For all sufficiently large \(d\),
the right-hand side is bounded below by some \(a_\epsilon>0\)
when \(k\leq(1-\epsilon)2d/\pi\), and above by \(-a_\epsilon\)
when \(k\geq(1+\epsilon)2d/\pi\). Thus, failure to detect in the
first regime or successful detection in the second requires
\(\mathcal D(\rho)\) to deviate from its mean by at least
\(a_\epsilon\sqrt{d/k}\). Applying
Eq.~\eqref{eq:pauli-concentration} with
\(t=(a_\epsilon/2)\sqrt{d/k}\) therefore yields
\begin{align}
\Pr[\mathcal D(\rho)\leq1]
&\leq Ce^{-c_\epsilon d},
&& k\leq(1-\epsilon)\tfrac{2d}{\pi},
\label{eq:pauli-below-threshold-tail}\\
\Pr[\mathcal D(\rho)>1]
&\leq Ce^{-c_\epsilon d},
&& k\geq(1+\epsilon)\tfrac{2d}{\pi},
\label{eq:pauli-above-threshold-tail}
\end{align}
where \(c_\epsilon>0\) depends only on \(\epsilon\).
This proves Theorem~\ref{thm:pauli-detection}.

\section{Proof of Theorem~\ref{coro:magic-detection}}
\label{app:magic-detection}

Let $\rho\sim\pi_{d,k}$, and fix observables $O_1,\ldots,O_M$ independently of $\rho$.
Certification requires that no stabilizer state be compatible with their expectation values.
Let $X=\rho-\mathbb{I}/d$, and let $E$ be the real span of the traceless parts of these observables.
Write $q=\dim E\leq M$, and let $P_E$ denote the Hilbert--Schmidt orthogonal projection onto $E$.
Set
\begin{equation}
r_0=\sqrt{\frac{a_\ast}{d(d-a_\ast)}}.
\end{equation}
Theorem~\ref{thm:purity-radius} gives $r_0B_2^m\subseteq K_n$.
Hence, if $\|P_E X\|_2\leq r_0$, then $\sigma=\mathbb{I}/d+P_E X$ belongs to $\operatorname{STAB}_n$.
Moreover, $\operatorname{Tr}(O_i\sigma)=\operatorname{Tr}(O_i\rho)$ for every $i$, since $X-P_E X$ is orthogonal to the traceless part of each observable.
Thus the protocol can certify magic only if $\|P_E X\|_2>r_0$.

The Haar second-moment identity after tracing out the environment gives
\begin{equation}
\mathbb{E}(\rho\otimes\rho)
=\frac{k\mathbb{I}_{d^2}+S}{d(dk+1)},
\end{equation}
where $S$ is the swap operator on $\mathbb{C}^d\otimes\mathbb{C}^d$.
So for every traceless Hermitian operator $A$ satisfying $\operatorname{Tr}(A^2)=1$,
\[
\mathbb{E}[\operatorname{Tr}(A\rho)^2]
=\frac{1}{d(dk+1)}.
\]
Summing over a Hilbert--Schmidt orthonormal basis of $E$ yields
\begin{equation}
\mathbb{E}\|P_E X\|_2^2=\frac{q}{d(dk+1)}.
\end{equation}
Let $p_{\mathrm{det}}$ denote the probability that the protocol certifies magic for $\rho\sim\pi_{d,k}$.
By Markov's inequality,
\begin{equation}
\begin{aligned}
p_{\mathrm{det}}
&\leq \Pr[\|P_E X\|_2>r_0]
\leq \frac{\mathbb{E}\|P_E X\|_2^2}{r_0^2}\\
&=\frac{q(d-a_\ast)}{a_\ast(dk+1)}
\leq \frac{M(d-a_\ast)}{a_\ast(dk+1)}.
\end{aligned}
\end{equation}
If $p_{\mathrm{det}}\geq p_0$ for a constant $p_0>0$ independent of $d$ and $k$, then
\begin{equation}
M\geq p_0a_\ast\frac{dk+1}{d-a_\ast}
=\Omega(k).
\end{equation}
This proves Theorem~\ref{coro:magic-detection}.

\section{Proof of Theorem~\ref{thm:facet-complexity}}
\label{app:facet-complexity}

By Theorem~\ref{thm:purity-radius}, $K_n$ contains a
Hilbert--Schmidt neighborhood of the origin. Hence $K_n$ is
full-dimensional and $0\in\operatorname{int}(K_n)$. Since translation
preserves the face lattice, polarity gives
\begin{equation}
F_d=\left|\operatorname{vert}(K_n^\circ)\right|.
\label{eq:facets-polar-vertices}
\end{equation}
For the lower bound, Eq.~\eqref{eq:few-vertex} gives $\operatorname{vrad}(\widetilde K_n)=\mathcal{O}(\log d/d)$.
Reverse Santal\'o, Eq.~\eqref{eq:reverse-santalo} then gives $\operatorname{vrad}(\widetilde K_n^\circ)=\Omega(d/\log d)$.
Since $K_n\subseteq\widetilde K_n$, polarity implies $\widetilde K_n^\circ\subseteq K_n^\circ$, and therefore
\begin{equation}
\operatorname{vrad}(K_n^\circ)
=
\Omega\!\left(\frac{d}{\log d}\right).
\label{eq:polar-vrad-facet}
\end{equation}

Theorem~\ref{thm:purity-radius} also gives $r_0B_2^m\subseteq K_n$ with $r_0=\sqrt{a_\ast/[d(d-a_\ast)]}$, hence $K_n^\circ\subseteq r_0^{-1}B_2^m$.
Define $P:=r_0K_n^\circ$.
Then $P\subseteq B_2^m$, $P$ has exactly $F_d$ vertices, and $\operatorname{vrad}(P)=\Omega(1/\log d)$.
Applying Eq.~\eqref{eq:few-vertex} to $P$ gives $\log(eF_d/m)=\Omega(m/\log^2d)$.
Since $m=d^2-1$,
\begin{equation}
F_d
\geq
\exp\!\left[
\Omega\!\left(\frac{d^2}{\log^2 d}\right)
\right].
\label{eq:facet-lower}
\end{equation}

For the upper bound, McMullen's theorem~\cite{mcmullen1970maximum} states that the cyclic polytope maximizes the number of facets among $m$-dimensional polytopes with $N_d$ vertices.
Since $m=d^2-1=2r+1$ is odd, the number of facets of cyclic polytope is $2\binom{N_d-r-1}{r}$.
Using $\binom{N}{r}\leq(eN/r)^r$, together with $m=\Theta(d^2)$ and $\log N_d=\Theta(\log^2d)$, gives
\begin{equation}
F_d
\leq
\exp\!\left[
\mathcal{O}\!\left(d^2\log^2 d\right)
\right].
\label{eq:facet-upper}
\end{equation}
Combining Eqs.~\eqref{eq:facet-lower} and~\eqref{eq:facet-upper} proves Theorem~\ref{thm:facet-complexity}.

\end{document}